\documentclass[12pt]{article}
\usepackage{amsmath}
\usepackage{amssymb}
\usepackage{hyperref}

\DeclareMathOperator{\bilat}{\!\! \stackrel{\leftrightarrow}{\partial}_{\!t} \!}

\begin{document}

\markboth{Pascu Gabriel}
{New Spherical Scalar Modes on the de Sitter Expanding Universe}


\title{New Spherical Scalar Modes on the de Sitter Expanding Universe}

\author{Pascu Gabriel \\
\small Faculty of Physics, West University of Timi\c{s}oara\\ 
\small Vasile P\^{a}rvan Avenue 4, Timi\c{s}oara, 300223, Rom\^{a}nia, EU\\
\small gpascu@physics.uvt.ro
}

\maketitle


\begin{abstract}
New spherical scalar modes on the expanding part of Sitter spacetime, eigenfunctions of a conserved Hamiltonian-like operator are found by solving the Klein-Gordon equation in the appropriate coordinate chart, with the help of a time evolution picture technique specially developed for spatially flat FLRW charts. Transition coefficients are computed between these modes and the rest of the scalar spherical and plane wave modes, either momentum or energy eigenfunctions on the spatially flat FLRW chart. \\\\

\end{abstract}


\newpage
\section{Introduction}

The de Sitter spacetime is the (one and only) positively curved maximal symmetric spacetime, in other words the positively curved spacetime that has the same number of Killing vectors as Minkowski spacetime. As such, is an ideal candidate for developing a quantum field theory on a curved but fixed background, covered with a certain coordinate chart. 
Following the steps of the canonical quantisation of the flat case, there has been constructed a quantum field theory on the spatially flat Friedmann-Lema\^{i}tre-Robertson-Walker (FLRW) chart. \cite{cotaescu2008quantum}

The Killing vectors of the manifold give rise to the generators of the representations of the isometry group carried by spaces of matter fields. \cite{cotaescu2000external} Therefore, the generators are operators related to the conserved quantities given by the Noether theorem, and from them one constructs the conserved one-particle operators of the corresponding quantum field theory.

The generators also commute with the operator of the field equation. In this respect, they or any other operator which commutes with that of the field equation are called conserved operators.
It is important to point out that a set of mode functions of a specified field equation of given spin on a spacetime (covered with a certain chart) is completely determined by the complete set of commutative operators (CSCO) $\{{\cal E},A,B,C\}$  chosen to label the modes. These must include the operator of the field equation ${\cal E}$, and some operators $A,B,C$, of which the modes are eigenfunctions, the number of which depends on the spacetime dimension and if the fields have spin. For scalar fields on a 3+1 dimensional  spacetime, 3 such operators are needed. If these operator are differential, then their eigenvalues can be thought of as separation constants for the Klein-Gordon equation. For fields with spin, an additional operator is required, which encodes in some way the spin. \cite{cotaescu2006remarks}

Though the scalar field is widely used in cosmology \cite{schrodinger1956expanding,tagirov1973consequences}, there hasn't yet been found evidence of a scalar elementary particle. Nevertheless, the free scalar field can be a good model for elucidating the inner workings of a field theory. \cite{candelas1975general,bunch1978quantum,pfautsch1982new}

A number of solutions for the Klein-Gordon equation in different charts on the de Sitter spacetime have been found, such as the static chart \cite{lohiya1978massless}, the hyperspherical one \cite{chernikov1968quantum}, or the spatially-flat FLRW one. \cite{nachtmann1967quantum} These three solutions are the best known, being treated even in classical texbooks. \cite{birrel1984quantum} 

There have been computed even Bogolyubov coefficients between the modes on the above mentioned charts. \cite{sato1994bogoliubov} Moreover, particle creation can be explained as emerging from non-null $\beta$-Bogolyubov coefficients, especially due to the thermal nature of the de Sitter horizon, and a number of studies have been conducted in this regard. \cite{allen1987massless,gibbons1977cosmological}

Other new modes can be defined either in new charts, either with new operators on the old charts. Recently, the theory of free scalar fields on de Sitter spacetime has also been discussed in Beltrami charts. \cite{wang2008classic} And a new hamiltonian-like operator has been on the spatially flat FLRW charts. \cite{cotaescu2008quantum} In fact, two spatially flat FLRW coordinate charts are needed in order to cover the whole of de Sitter space. These two charts don't overlap, one of them describes an expanding spacetime to which we refer as "the de Sitter expanding universe", and the other a contracting one.
The mode solutions on the spatially-flat FLRW chart are also momentum (or momentum squared) eingenvectors (called "momentum basis modes", either plane-wave modes- eigenfunctions of $\{{\cal E}_{KG},\vec{P}\}$\cite{nachtmann1967quantum}, or spherical ones- eigenfunctions of $\{{\cal E}_{KG},\vec{P}^2,\vec{L}^2,L_z\}$, as used in \cite{sato1994bogoliubov}). 

Recently, Cot\u{a}escu et. al \cite{cotaescu2008quantum} managed to find, through a technique involving time evolution pictures on spatially flat FLRW manifolds, scalar plane wave solutions that are eigenfunctions of a newly-defined conserved operator, a modified Hamiltonian $H$ (modes of which are called "energy basis" plane wave modes, and are eigenfunctions of $\{{\cal E}_{KG},H,{\cal N}_i\}$).

They also argued that, due to the uncertainty between the energy and momentum on de Sitter spacetime, both momentum basis and energy basis quantum modes are required in order to build a complete de Sitter quantum field theory, for which some applications have already been worked out.\cite{crucean2010amplitude,crucean2007coulomb}

Still, the spherical energy basis scalar quantum modes (eigenfunctions of $\{{\cal E}_{KG},H,\vec{L}^2,L_z\}$) remained unsolved. This paper is devoted to finding the analytical form of these scalar modes, on the spatially flat de Sitter FLRW chart, via the Schr\"{o}dinger picture (SP) of the time evolution technique developed specially for spatially flat FLRW spacetimes. \cite{cotaescu2007schrodinger}

In the second section, we make a brief review of the known results, with the formalism used afterwards to derive the new results. This is done not just to accommodate the reader with the already known solutions or used conventions, but is necessary in order to understand both the similarities and mainly the differences between the momentum basis quantum modes and the corresponding energy basis ones. The second section concludes with expressing the planar energy basis modes in a new form, as a sum of two hypergeometric functions.
The third section contains entirely original work. It consists of the first derivation of the new spherical scalar energy-basis modes, and a way of determining them through a different method of the one used to obtain the plane-wave correspondents.\cite{cotaescu2008quantum} 
The paper concludes with the fourth section where the transition coefficients between all the aforementioned modes are briefly computed, showing that the energy basis modes are indeed different from the previously known momentum basis ones.

\section{Previous solutions on the spatially flat FLRW chart}
\subsection{Momentum basis plane waves}
The Klein-Gordon equation, on the 3+1 dimensional de Sitter spacetime, in the spatially flat FLRW chart with line element $ds^2=dt^2-e^{2\omega t}d\vec{x}\cdot d\vec{x}$, parametrised by (spatial) Cartesian coordinates $\{t,\vec{x}\}$ takes the form
\begin{equation}\label{KG_tx}
\left( \partial_t^2+ 3\omega \partial_t- e^{-2\omega t}\Delta_{x,y,z}+ m^2 \right)\Phi(t,\vec{x})=0.
\end{equation}

The first quantum modes ever written on the de Sitter spacetime, were the eigenfunctions of the momentum operator, in this chart.\cite{nachtmann1967quantum} By promptly solving the three eigenvalue equations $P_i \Phi(t,\vec{x})=p_i \Phi(t,\vec{x})$, one obtains that the fundamental solution of the Klein-Gordon equation is in this case separable as $f_{\vec{p}}(t,\vec{x})=Nf_p(t) e^{i\vec{p}\cdot \vec{x}}$, which leads to the equation for the temporal part:

\begin{equation}
\frac{d^2}{dt^2} f_p(t)+ 3\omega \frac{d}{dt}f_p(t)+ (p^2 e^{-2\omega t}+m^2)f_p(t)=0,
\end{equation}
which has the solution: 
\begin{equation}\label{f_temp}
f_p(t)=Ce^{\frac{3\omega t}{2}} H^{(1)}_\nu \left( \frac{p}{\omega} e^{-\omega t} \right).
\end{equation}
with $\nu=i \sqrt{\mu^2-9/4}$.

It is worth pointing out that the choice of the function $H_\nu^{(1)}$ as the solution of the Bessel equation is due to the fact that one chooses to work with progressive waves, rather than regressive ones ($H_\nu^{(2)}$- corresponding to $f^*$). 

The fundamental solutions of the equation can be normalisation with the help of the scalar product expressed in the Cartesian chart as

\begin{equation}\label{scprod_x}
\langle f(t,\vec{x}) ,f^\prime (t,\vec{x})  \rangle=i e^{3\omega t} \int_{\mathbf{R}^3} d^3x f^* (t,\vec{x}) \bilat f^\prime (t,\vec{x}),
\end{equation}
which involves the Wronskian of two Hankel functions
\begin{equation}\label{Hwronsk}
\left( H^{(1)}_\nu \left(\frac{p}{\omega}e^{-\omega t}\right) \right)^* \bilat H^{(1)}_\nu\left(\frac{p}{\omega} e^{-\omega t}\right)=-\frac{4i\omega}{\pi} e^{ -i \pi \nu \Theta(\mu-3/2)}.
\end{equation} 
Here, two cases are included: when $\nu$ is real and $\mu \leqslant 3/2$, and when $\nu$ is pure imaginary and $\mu > 3/2$, where an extra factor $e^{-i\pi \nu}$ appears, $\Theta$ being just the Heaviside step function used to encompass both cases. With these in mind, the normalised modes have the following expression:
\begin{equation}\label{mbpw}
f_{\vec{p}}(t,\vec{x}) = \frac{1}{2}\sqrt{\frac{\pi}{\omega}} \frac{1}{(2\pi)^{3/2}} e^{-\frac{3\omega t}{2}} e^{ \frac{i \pi \nu}{2} \Theta(\mu-3/2)} H^{(1)}_\nu \left( \frac{p}{\omega} e^{-\omega t} \right) e^{i\vec{p}\cdot\vec{x}}.
\end{equation}

\subsection{Momentum basis spherical waves}
The Klein-Gordon equation on this chart (\ref{KG_tx}) can be put in spherical coordinates. 
The momentum basis for spherical waves is given by the CSCO $\{{\cal E}_{KG},\vec{P}^2,\vec{L}^2,L_z\}$, so its fundamental solution must be of the form 
\begin{equation}
f_{p,l,m_l}(t,r,\theta,\phi)=N f_p(t) R_{p,l}(r) Y_{l, m_l}(\theta,\phi), 
\end{equation}
where the radial part is given by the eigenvalue equation $\vec{P}^2 R_{p,l}(r)=p^2 R_{p,l}(r)$. Using the Laplacian in spherical coordinates leads to a spherical Bessel differential equation:
\begin{equation}
\frac{d^2}{dr^2}R_{p,l}(r)+\frac{2}{r}\frac{d}{dr}R_{p,l}(r)- \left( \frac{l(l+1)}{r^2}-p^2 \right) R_{p,l}(r)=0.
\end{equation}
The solution can be written as $R_{p,l}(r)=Cj_l(rp)=C^\prime \frac{1}{\sqrt{r}} J_{l+\frac{1}{2}}(rp)$, where the temporal part $f_p(t)$ is the same as for the plane waves (\ref{f_temp}), such that the normalised modes are in this case
\begin{equation}\label{mbsw}
f_{p,l,m_l}(t,r,\theta,\phi) = \frac{1}{2}\sqrt{\frac{p\pi}{\omega}} e^{-\frac{3\omega t}{2}} e^{ \frac{i \pi \nu}{2} \Theta(\mu-3/2)} \frac{1}{\sqrt{r}} H^{(1)}_\nu \left( \frac{p}{\omega} e^{-\omega t} \right) J_{l+1/2}(pr) Y_{l,m_l}(\theta,\phi).
\end{equation}
by using the scalar product expressed in spherical coordinates, which reads:
\begin{equation}\label{scprod_r}
\langle f(t,r,\theta,\phi) ,f^\prime (t,r,\theta,\phi)  \rangle=i e^{3\omega t} \int_0^\infty dr r^2  \int_{\mathbf{S}^2} d\Omega f^* (t,r,\theta,\phi) \bilat f^\prime (t,r,\theta,\phi).
\end{equation}

It should be noted that the spherical momentum modes can also be obtained from the plane wave ones, by performing a Rayleigh expansion- expanding the plane wave into a sum of spherical waves: 
\begin{equation}\label{rayleigh}
e^{i\vec{q}\cdot\vec{x}}=(2\pi)^{\frac{3}{2}} \frac{1}{\sqrt{qr}}\sum\limits_{l=0}^\infty \sum\limits_{m_l=-l}^l i^l J_{l+\frac{1}{2}}(qr) Y_{l m_l}(\theta,\phi) Y^*_{l m_l}(\theta_{\vec{q}},\phi_{\vec{q}}).
\end{equation}

\subsection{Energy basis plane waves}

Cot\u{a}escu et al. found the scalar modes that are eigenfunctions of a new Hamiltonian. \cite{cotaescu2008quantum} Since the $-i\partial_t$ is not conserved in the sense that is doesn't commute with the field equation operator, the energy basis is defined as being given by the CSCO $\{ {\cal E}_{KG},H,{\cal N}_i\}$, where $H= -i\partial_t +\omega x^i\partial_i$ is a modified conserved Hamiltonian, and ${\cal N}_i$ are any two of the three non-differential operators which encode the momentum vector's direction.  

Passing from the natural picture (NP) to the Schr\"{o}dinger picture (SP), according to the rigorous procedure fully described in \cite{cotaescu2008quantum}, the Klein-Gordon equation of the natural picture (\ref{KG_tx}) becomes in the Schr\"{o}dinger picture, in the same chart:
\begin{equation} \label{KG_tx_SP}
\left[ (\partial_t+ \omega x^i \partial_i)^2+ 3\omega (\partial_t+ \omega x^i \partial_i)- \Delta_{x,y,z}+ m^2 \right]\Phi_S(t,\vec{x})=0.
\end{equation}

Expanding the field as $\Phi_S(x)=\Phi_S^{(+)}(x)+\Phi_S^{(-)}(x)$ with
\begin{equation}
\Phi_S^{(\pm)}(x)=\int_0^\infty dE \int d^3q \, \, \hat{\Phi}_S^{(\pm)}(E,\vec{q}) e^{\mp i(Et-\vec{q}\cdot\vec{x})},
\end{equation}
one arrives to the Klein-Gordon equation in momentum representation, in the SP:
\begin{equation}
\left[ \left( \pm iE+ \omega (q^i \partial_{q_i} +3) \right)^2 -3\omega \left( \pm iE +\omega (q^i \partial_{q_i} +3) \right) + \vec{q}^2 +m^2 \right] \hat{\Phi}_S^{(\pm)}(E,\vec{q})=0, 
\end{equation}
where for example for the positive frequency part $\hat{\Phi}^{(+)}_S(E,\vec{q})=h_S(E,q) \, a(E,\vec{n})$, and the function $h_S$ being of radial type, it satisfies the equation
\begin{equation}
\left[ \left( \pm iE+ \omega (q \partial_{q} +3) \right)^2 -3\omega \left( \pm iE +\omega (q \partial_{q} +3) \right) + q^2 +m^2 \right] h_S^{(\pm)}(E,q)=0, 
\end{equation}
which can be shown to have solutions of the form: $h_S(\epsilon,s)=C s^{-i\epsilon -3/2} H_\nu^{(1)}(s)$, where the variables $\epsilon=E/\omega$,$s=q/\omega$ and $\mu=m/\omega$ have been rescaled for simplicity. Again, the choice $H^{(1)}_\nu$ as the solution of the Bessel equation is made due to the convention that we express $f$ using progressive waves.

From the desired expansion of the scalar field in SP, the expression for the mode functions (in SP) can be read off as:
\begin{equation}
f^S_{E,\vec{n}}(t,\vec{x})= N e^{-iEt} \int_0^\infty ds \sqrt{s}s^{-i\epsilon} H_\nu^{(1)}(s),
\end{equation}
and by passing back to the one in NP, followed by normalization the energy basis modes (in integral representation) are obtained:
\begin{equation}\label{ebpw}
f_{E,\vec{n}}(t,\vec{x}) = \frac{1}{2}\sqrt{\frac{\omega}{2}} \frac{1}{(2\pi)^{3/2}} e^{-\frac{3\omega t}{2}} e^{ \frac{i \pi \nu}{2} \Theta(\mu-3/2)} \int_0^\infty ds \sqrt{s} s^{-i\epsilon} H^{(1)}_\nu (s e^{-\omega t}) e^{i\omega s \vec{n}\cdot\vec{x}}.
\end{equation}

This is the form in which the plane-wave energy basis scalar modes were first given.\cite{cotaescu2008quantum} The authors searched for the plane waves which are eigenfunctions of a CSCO that included non-differential operators ${\cal N}_i$. Due to this, one could not simply solve the equation by using the extra eigenvalue equations, and had to resort to writing the equation in the momentum representation. However, as we shall see, this is not the case for the energy basis spherical waves.

Furthermore, the integral in the expression of the modes can be solved, giving a Gauss hypergeometric function. 
\begin{align}
f_{E,\vec{n}}(t,\vec{x}) = \frac{\sqrt{\omega}}{8\pi} e^{-\frac{3\omega t}{2}} e^{ \frac{i \pi \nu}{2} \Theta(\mu-3/2)} \frac{(-2i e^{-\omega t})^\nu}{(-i\omega \vec{n}\cdot\vec{x}-ie^{-\omega t})^{-i\epsilon+3/2+\nu}} \notag \\
 \times {}_2 F_1 \left(-i\epsilon+\frac{3}{2}+\nu,\nu+\frac{1}{2};-i\epsilon+2; \frac{\omega \vec{n}\cdot\vec{x} e^{\omega t}-1}{\omega \vec{n}\cdot\vec{x} e^{\omega t}+1}  \right).
\end{align}

In this paper, we present a new explicit form for the above expression of the energy basis scalar modes, which can be written in terms of other hypergeometric functions\cite{gr} as:
\begin{align}
f_{E,\vec{n}}(t,\vec{x}) =\frac{\sqrt{\omega}}{(2\pi)^{5/2}} e^{-\frac{E\pi}{2\omega}} e^{ -iEt}  {\bigg [} \Gamma(\sigma_+)\Gamma(\sigma_-) {}_2 F_1 \left(\sigma_+,\sigma_-;\frac{1}{2};\omega^2 (\vec{n}\cdot\vec{x})^2 e^{2\omega t}  \right) \notag \\
- 2\omega \vec{n}\cdot\vec{x} e^{\omega t} \Gamma\left(\sigma_++\frac{1}{2}\right)\Gamma\left(\sigma_-+\frac{1}{2}\right)  {}_2 F_1 \left(\sigma_++\frac{1}{2},\sigma_-+\frac{1}{2};\frac{3}{2};\omega^2 (\vec{n}\cdot\vec{x})^2 e^{2\omega t} \right) {\bigg ]},
\end{align}
where $\sigma_\pm=\frac{3}{4}-\frac{i\epsilon}{2}\pm\frac{\nu}{2}$. This is the form which can be compared with the new results that will be deduced in the next section.

\section{Energy basis spherical waves on the spatially flat FLRW chart}

We now proceed to the main original part of this work- finding the spherical energy basis quantum modes. These are defined as eigenfunctions of the CSCO $\{ {\cal E}_{KG},H,\vec{L}^2,L_z\}$. Unlike the plane wave energy basis modes, the spherical ones being eigenfunctions of differential operators, can be found by solving the Klein-Gordon equation, in the spherical FLRW chart, using the associated eigenvalue equations. One of the main points of the paper is showing how these modes can be found, by solving the equation in this way.

In the Klein-Gordon equation in SP (\ref{KG_tx_SP}), we pass from the Cartesian coordinates $\{t,\vec{x}\}$ to spherical coordinates $\{t,r,\theta,\phi\}$, so the equation changes to:
\begin{equation}
\left( (\partial_t+ \omega r \partial_r)^2+ 3\omega (\partial_t+ \omega r \partial_r)- \partial_r^2- \frac{2}{r}\partial_r- \frac{\Delta_{\theta,\phi}}{r^2}+ m^2 \right)\Phi_S(t,r,\theta,\phi)=0.
\end{equation}

The "energy basis spherical" quantum modes we wish to find are solutions of this equation, but (as their name implies) they are also eigenfunctions of the following operators: $H,\vec{L}^2,L_z$.
By promptly solving the eigenvalues equations, while being in the SP, for the CSCO $H=-i\partial_t, \vec{L}^2= -\Delta_{\theta,\phi}, L_z= -\Delta_\phi$, it is straightforward to show that the solution of the equation must include the factor $e^{-iEt}$, and the spherical harmonics $Y_{l,m_l}(\theta,\phi)$, such that the fundamental solutions are separable as:
\begin{equation}
f^S_{E,l,m_l}(t,r,\theta,\phi)=N e^{-iEt} R^S_{E,l}(r) Y_{l,m_l}(\theta,\phi).
\end{equation}

It is of note that the separation of variables occurs only here in the SP, and only thanks to the use of this time evolution technique. \cite{cotaescu2008quantum} We stress again that in SP, $H=-i\partial_t$ is conserved, while in the NP only the modified Hamiltionian $H=-i\partial_t-\omega x^i \partial_i$ is conserved. By inputting the above solution, and making use of the chosen operator eigenvalue equations, the Klein-Gordon equation becomes an ordinary differential equation for the radial part:

\begin{align}
{\bigg [} (\omega^2 r^2-1)\frac{d^2}{dr^2}+ \left( 4\omega^2 r - 2i\omega E r -\frac{2}{r} \right) \frac{d}{dr} \notag \\
-E^2 - 3i\omega E+ \frac{l(l+1)}{r^2}+ m^2 {\bigg ]} R^S_{E,l}(r)=0.
\end{align}

Changing the variable from $r$ to $z=\omega^2 r^2$, and dividing by $ -4\omega^2$, the equation becomes: 
\begin{equation}
\left[z(1-z) \frac{d^2}{dz^2}+\left(-\frac{5z}{2}+\frac{3}{2}+i\epsilon z \right)\frac{d}{dz}+ \left( \frac{\epsilon^2}{4}+\frac{3i\epsilon}{4}- \frac{l(l+1)}{4z}- \frac{\mu^2}{4} \right)  \right]R^S_{E,l}(z)=0,
\end{equation}
where for brevity $\epsilon=E/\omega$ and $\mu=m/\omega$.

Making an ansatz for $R_S(z)$ as $R^S_{E,l}(z)=z^{\frac{l}{2}} S^S_{E,l}(z)$, we obtain an equation which has the form of the Gauss hypergeometric equation \cite{nist}
\begin{align}
{\bigg [} z(1-z) \frac{d^2}{dz^2}+\left(l+\frac{3}{2} -\left(\frac{5}{2}-i\epsilon\right)z \right)\frac{d}{dz} \notag \\
+\left( \frac{l^2}{4}-\frac{3l}{4}+\frac{3i\epsilon}{4}+ \frac{i\epsilon l}{2}+\frac{\epsilon^2}{4}- \frac{\mu^2}{4} \right)  {\bigg ]} S^S_{E,l}(z)=0.
\end{align}

Discarding the solution that is singular at the origin, we remain only with:
\begin{equation}
S^S_{E,l}(z)={}_2 F_1\left(\sigma_+-\frac{l}{2},\sigma_--\frac{l}{2};l+\frac{3}{2};z\right),
\end{equation}
where again $\sigma_\pm=\frac{3}{4}-\frac{i\epsilon}{2}\pm\frac{\nu}{2}$ and $\nu=i\sqrt{\mu^2-\frac{9}{4}}$.

Bearing in mind all of the above, and by passing back from the variable $z$ to $r$:
\begin{equation}
f^S_{E,l,m_l}(t,r,\theta,\phi)=N e^{-iEt} (\omega r)^l {}_2 F_1\left(\sigma_+-\frac{l}{2},\sigma_--\frac{l}{2};l+\frac{3}{2};\omega^2 r^2\right) Y_{l,m_l}(\theta,\phi),
\end{equation}
and then from SP to NP, we obtain the following solution to the Klein-Gordon equation:
\begin{equation}
f_{E,l,m_l}(t,r,\theta,\phi)=N e^{-iEt} (\omega r e^{\omega t})^l {}_2 F_1\left(\sigma_+-\frac{l}{2},\sigma_--\frac{l}{2};l+\frac{3}{2};\omega^2 r^2 e^{2\omega t}\right) Y_{l,m_l}(\theta,\phi).
\end{equation}

Next, these quantum modes will be normalised. In order to do that, part of the solution must be written as a Hankel transform of a certain function:
\begin {equation}
f_{E,l,m_l}(t,r,\theta,\phi)=N e^{-iEt} (\omega e^{\omega t})^l \frac{1}{\sqrt{r}}  Y_{l,m_l}(\theta,\phi) \mathcal{H}_{\lambda}[g(\omega r)],
\end{equation}
where
\begin {equation}
\mathcal{H}_{\lambda}[g(\omega r)]=r^{l+\frac{1}{2}} {}_2 F_1\left(\sigma_+-\frac{l}{2},\sigma_--\frac{l}{2};l+\frac{3}{2};\omega^2 r^2 e^{2\omega t}\right).
\end{equation}

According to \cite{gr}, this is the Hankel transform of order $\lambda=l+\frac{1}{2}$ of:
\begin {equation}
g(s)= \omega^2 \int_0^\infty r^{l+\frac{3}{2}} {}_2 F_1\left(\sigma_+-\frac{l}{2},\sigma_--\frac{l}{2};l+\frac{3}{2};\omega^2 r^2 e^{2\omega t}\right) J_{l+\frac{1}{2}}(\omega r s) dr,
\end{equation}
which can be evaluated as
\begin {align}
g(s)= 2^{i\epsilon} i^{-\frac{1}{2}-l+i\epsilon} e^{iEt-\omega t \left(l+\frac{3}{2}\right)} \omega^{-l-\frac{1}{2}} \pi e^{\frac{i\pi\nu}{2}} \notag \\
\times \frac{\Gamma \left(l+ \frac{3}{2}\right)}{\Gamma\left(\sigma_+-\frac{l}{2}\right)\Gamma\left(\sigma_--\frac{l}{2}\right)} s^{-i\epsilon-1} H^{(1)}_\nu(se^{-\omega t}),
\end{align}
such that 
\begin {align}
\mathcal{H}_{\lambda}[g(\omega r)]=2^{i\epsilon} i^{-\frac{1}{2}-l} e^{-\frac{\epsilon \pi}{2}} e^{iEt-\omega t \left(l+\frac{3}{2}\right)} \omega^{-l-\frac{1}{2}} \pi e^{\frac{i\pi\nu}{2}} \notag \\
\times \frac{\Gamma \left(l+ \frac{3}{2}\right)}{\Gamma\left(\sigma_+-\frac{l}{2}\right)\Gamma\left(\sigma_--\frac{l}{2}\right)} \int_0^\infty s^{-i\epsilon} H^{(1)}_\nu(se^{-\omega t}) J_{l+\frac{1}{2}}(\omega rs)ds.
\end{align}

As a result of this, the quantum modes in integral representation can be written as:
\begin {align}
f_{E,l,m_l}(t,r,\theta,\phi)&=N 2^{i\epsilon} i^{-\frac{1}{2}-l} \omega^{\frac{3}{2}} e^{-\frac{\epsilon \pi}{2}} \pi e^{-\frac{3\omega t}{2}}  e^{\frac{i\pi\nu}{2}} \frac{\Gamma \left(l+\frac{3}{2}\right)}{\Gamma\left(\sigma_+-\frac{l}{2}\right)\Gamma\left(\sigma_--\frac{l}{2}\right)} \notag \\
\times Y_{l,m_l} &(\theta,\phi) \frac{1}{\sqrt{r}} \int_0^\infty s^{-i\epsilon} H^{(1)}_\nu(se^{-\omega t}) J_{l+\frac{1}{2}}(\omega rs)ds.
\end{align}

Applying the scalar product, written in spherical coordinates (\ref{scprod_r}), and taking note of the closure relations and the Wronskian of the Hankel functions (\ref{Hwronsk}), the normalisation constant is: 
\begin{equation}
N=\frac{e^{\frac{\epsilon \pi}{2}}}{2\sqrt{2}\pi \omega^{\frac{3}{2}}} \frac{\Gamma\left(\sigma_+-\frac{l}{2}\right)\Gamma\left(\sigma_--\frac{l}{2}\right)}{\Gamma \left(l+\frac{3}{2}\right)},
\end{equation}
such that the modes have the simple expression in the integral form (dropping the phase factors):
\begin{align}\label{ebsw}
f_{E,l,m_l}(t,r,\theta,\phi)=\frac{1}{2\sqrt{2}} e^{-\frac{3\omega t}{2}}  e^{\frac{i\pi\nu}{2}\Theta \left(\mu-\frac{3}{2}\right)}   Y_{l,m_l}(\theta,\phi) \notag\\
\times \frac{1}{\sqrt{r}} \int_0^\infty s^{-i\epsilon} H^{(1)}_\nu(se^{-\omega t}) J_{l+\frac{1}{2}}(\omega rs)ds,
\end{align}
or the full, normalized explicit expression
\begin{align}\label{ebsw_full}
f_{E,l,m_l}(t,r,\theta,\phi)= \frac{1}{2\sqrt{2}\pi \omega^\frac{3}{2}}\frac{\Gamma\left(\sigma_+-\frac{l}{2}\right)\Gamma\left(\sigma_--\frac{l}{2}\right)}{\Gamma \left(l+\frac{3}{2}\right)} e^{-iEt} (\omega r e^{\omega t})^l \notag \\
\times {}_2 F_1\left(\sigma_+-\frac{l}{2},\sigma_--\frac{l}{2};l+\frac{3}{2};\omega^2 r^2 e^{2\omega t}\right) Y_{l,m_l}(\theta,\phi).
\end{align}

The completeness relation for these mode functions can be verified with their integral form. 
 
Alternatively, one can use the Rayleigh expansion (\ref{rayleigh}) on the plane wave energy basis normalised modes (\ref{ebpw}) to obtain the corresponding spherical ones (\ref{ebsw}), which ensures the agreement between the above approach and the method of \cite{cotaescu2008quantum}.

These modes are important because they can be used as scalar solutions of given energy in scattering problems like partial wave analysis on de Sitter spacetime, where for instance an incoming plane wave of either definite momentum (\ref{mbpw}) or definite energy of the form (\ref{ebpw}) scatters off a central potential, resulting in a spherical wave of definite energy (\ref{ebsw}). This could shed light on the problem of correctly defining the scattering cross section on de Sitter spacetime. 

\section{Transition coefficients}
One can compute the scalar products between any of the quantum modes (\ref{mbpw}),(\ref{mbsw}),(\ref{ebpw}) and (\ref{ebsw}). For the scalar product between plane and spherical waves, it is necessary to expand the plane ones into spherical waves, via the Rayleigh expansion formula, in order to be able to use the scalar product in spherical coordinates (\ref{scprod_r}).

These transition coefficients have the following expressions:
\begin{align}
\langle f_{\vec{p}}, f_{E,\vec{n}} \rangle &= \frac{p^{-\frac{3}{2}}}{\sqrt{2\pi\omega}} \delta^2(\vec{n}-\vec{n}_p) e^{-i\frac{E}{\omega}\ln \frac{p}{\omega}}, \label{fsp}\\
\langle f_{\vec{p}}, f_{p,l,m_l} \rangle &= \frac{1}{p} \delta(p-|\vec{p}|) (-i)^l Y_{l,m_l}(\theta_{\vec{p}},\phi_{\vec{p}}),  \label{sp2}\\
\langle f_{\vec{p}}, f_{E,l,m_l} \rangle &= \frac{p^{-\frac{3}{2}}}{\sqrt{2\pi\omega}} (-i)^l Y_{l,m_l}(\theta_{\vec{p}},\phi_{\vec{p}}) e^{-i\frac{E}{\omega}\ln \frac{p}{\omega}}, \label{sp3}\\
\langle f_{E,\vec{n}}, f_{p,l,m_l} \rangle &= \frac{p^{-\frac{1}{2}}}{\sqrt{2\pi\omega}} (-i)^l Y_{l,m_l}(\theta_{\vec{n}},\phi_{\vec{n}}) e^{i\frac{E}{\omega}\ln \frac{p}{\omega}},\label{sp4}\\
\langle f_{E,\vec{n}}, f_{E^\prime,l,m_l} \rangle &= \delta(E-E^\prime) (-i)^l Y_{l,m_l}(\theta_{\vec{n}},\phi_{\vec{n}}), \label{sp5}\\
\langle f_{p,l,m_l}, f_{E,l^\prime,m_l^\prime} \rangle &= \frac{p^{-\frac{1}{2}}}{\sqrt{2\pi\omega}} \delta_{l,l^\prime} \delta_{m_l,m_l^\prime} e^{-i\frac{E}{\omega}\ln \frac{p}{\omega}}, \label{sp6}\\
\end{align}
while their converses satisfy:
\begin{equation}
\langle f_{2}, f_{1} \rangle = \langle f_{1}, f_{2} \rangle ^*.
\end{equation}

All these transition coefficients can be thought of as $\alpha$-Bogolyubov coefficients. However, the $\beta$-Bogolyubov coefficients between any of the discussed mode functions cancel out, thanks to the fact that the Wronskian of two Hankel functions of the same kind is null.

\begin{equation}
\langle f_{1}, f_{2}^* \rangle = 0.
\end{equation}

This means there is no mixing of positive and negative frequencies.

\section{Conclusions}

Scalar spherical quantum modes on the de Sitter expanding universe, that are eigenfunctions of a modified Hamiltonian operator have been computed, completing the free scalar field mode analysis on de Sitter spacetime in the spatially flat FLRW chart. \cite{cotaescu2008quantum} Our scalar modes, in their hypergeometric form remind of the ones found long ago on the static chart. \cite{lohiya1978massless} An explicit alternate form of the plane-wave energy basis quantum modes, in terms of two hypergeometric functions was also given, that presents similarities with the one found for the corresponding spherical modes. 
However, it is the integral form in which these mode functions are more useful, since they are expressed in terms of more suitable functions, and one can use them for orthnormalization, show completion or compute transition coefficients with ease.
The evaluated transition coefficients between both planar and spherical waves, in momentum and energy bases show that there is no positive-negative frequency mixing between any of the discussed mode functions. 

\section*{Acknowledgements}
\indent The author would like to thank Professor Ion I. Cot\u{a}escu for the fruitful discussions that resulted in invaluable suggestions from which this work has benefited, and also to dr. Cosmin Crucean for useful suggestions that helped improve this paper.


\end{document}